\documentclass[aps,PRL,twocolumn,superscriptaddress,floats]{revtex4}
\usepackage{amssymb}
\usepackage{graphicx}

\begin{document}

\title{Strong correlation and massive spectral-weight redistribution induced spin density wave in $\alpha$-Fe$_{1.06}$Te}

\author{Y. Zhang}
\author{F. Chen}
\author{C. He}
\author{L. X. Yang}
\author{B. P. Xie}

\affiliation{Department of Physics, Surface Physics Laboratory
(National Key Laboratory), and Advanced Materials Laboratory, Fudan
University, Shanghai 200433, People's Republic of China}

\author{Y. L. Xie}
\author{X. H. Chen}

\affiliation{Department of Physics,  University of Science and
Technology of China, Hefei, Anhui 230027, People's Republic of
China}

\author{Minghu Fang}

\affiliation{Department of Physics, Zhejiang University, Hangzhou
310027, China}

\author{M. Arita}
\author{K. Shimada}
\author{H. Namatame}
\author{M. Taniguchi}

\affiliation{Hiroshima Synchrotron Radiation Center and Graduate
School of Science, Hiroshima University, Hiroshima 739-8526, Japan.}

\author{J. P. Hu}

\affiliation{Department of Physics, Purdue University, West
Lafayette, Indiana 47907, USA}

\author{D. L. Feng}
\email{dlfeng@fudan.edu.cn}

\affiliation{Department of Physics, Surface Physics Laboratory
(National Key Laboratory), and Advanced Materials Laboratory, Fudan
University, Shanghai 200433, People's Republic of China}

\date{\today}

\begin{abstract}
The electronic structure of $\alpha$-Fe$_{1.06}$Te is studied with
angle-resolved photoemission spectroscopy. We show that there is
substantial spectral weight around $\Gamma$ and $X$,  and lineshapes
are intrinsically incoherent in the paramagnetic state. The magnetic
transition is characterized by a massive spectral-weight transfer
over an energy range as large as the band width, which even exhibits
a hysteresis loop that marks the strong first order transition.
Coherent quasiparticles emerge in the magnetically ordered state due
to decreased spin fluctuations, which account for the change of
transport properties from insulating behavior to metallic behavior.
Our observation demonstrates that Fe$_{1.06}$Te distinguishes itself
from other iron-based systems with more local characters and much
stronger interactions among different degrees of freedom, and how a
spin density wave is formed in the presence of strong correlation.
\end{abstract}

\pacs{74.25.Jb,74.70.Xa,79.60.-i,71.20.-b}

\maketitle

\begin{figure}[t]
\includegraphics[width=8cm]{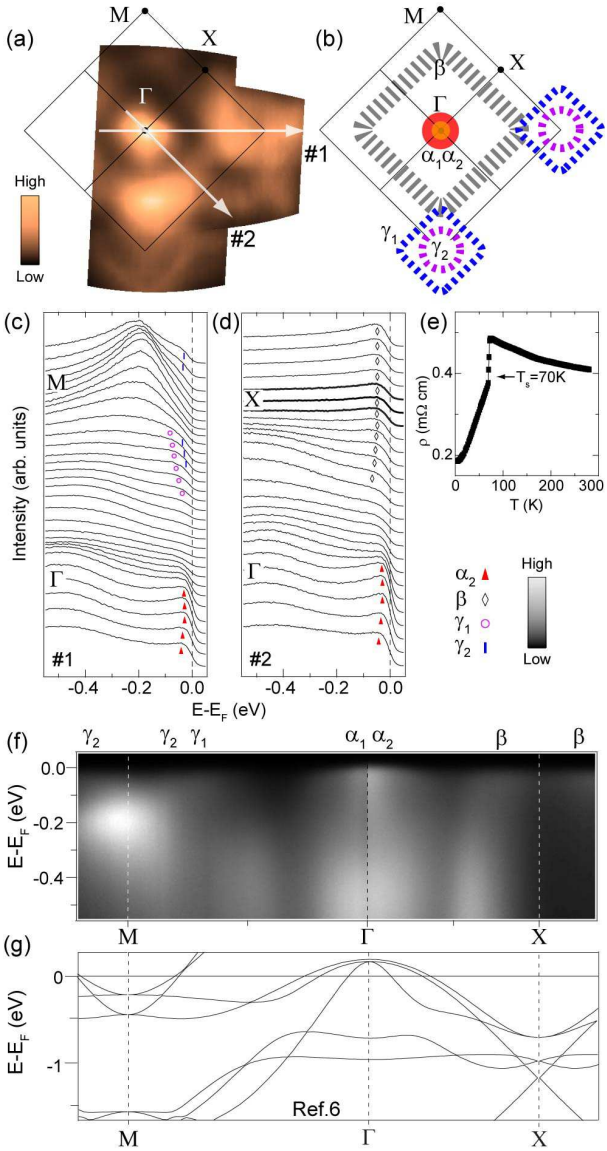}
\caption{(Color online) Paramagnetic state electronic structure of
Fe$_{1.06}$Te measured at 135~K. (a) Photoemission intensity
distribution integrated over the energy window of $[E_F-15~meV,
E_F+15~meV]$. (b) The spectral-weight distribution around $E_F$,
which are labeled by the filled orange circle ($\alpha_1$), the red
circle ($\alpha_2$), the dashed black square ($\beta$), the dashed
blue squares ($\gamma_1$), and dashed purple squares ($\gamma_2$).
(c) and (d) The energy distribution curves (EDCs) along cut~\#1 and
\#2 respectively. (e) The temperature dependence of the resistivity
of Fe$_{1.06}$Te. (f) The photoemission intensities along the
$M-\Gamma-X$ high symmetry lines, and (g) the corresponding band
structure based on DFT calculations. \cite{DJSingh1}.}
\label{normal}
\end{figure}

The discovery of iron-based high-temperature superconductors
(Fe-HTSCs) has generated great interests \cite{Hosono2008}. So far,
two classes of Fe-HTSC  have been discovered. They are iron
pnictides, \textit{e.g.}, SmO$_{1-x}$F$_x$FeAs or
Ba$_{1-x}$K$_{x}$Fe$_{2}$As$_{2}$ \cite{XHChen, Johrendt1}, and iron
chalcogenides, \textit{e.g.}, Fe$_{1+y}$Te$_{1-x}$Se$_{x}$
\cite{Wu2}. Although, both classes of materials share many common
aspects, such as similarly high maximal superconducting transition
temperature ($T_c$) (Fe$_{1+y}$Se possesses a $T_c$ of 37~K under
hydrostatic pressure of 7~GPa \cite{Prassides}) and similar band
structures from density-functional theory (DFT) calculations
\cite{DJSingh1,DJSingh122}. However, their parent compounds exhibit
quite different spin density wave (SDW) states. A collinear
commensurate antiferromagnetic order has been identified for the
pnictides \cite{PCDai1111,XHChen122}, while a bicollinear and
45-degree rotated antiferromagnetic order was identified for
Fe$_{1+y}$Te \cite{MaoN,PCDai}. Furthermore, the transport
properties of Fe$_{1+y}$Te respond abruptly to the first order
magnetic/structural transition. In the paramagnetic state, it shows
insulator-like resistivity  [Fig.~\ref{normal}(e)], and optical
conductivity without a Drude peak, while the resistivity becomes
metallic-like, and a Drude peak emerges in the SDW state
\cite{NLWang, MaoT}.

Like the cuprates, the nature of magnetic order and spin
fluctuations in Fe-HTSC are most likely crucial for its
superconductivity. Yet the origin of the magnetic ordering in iron
pnictides/chacogenides is still under heated debate. For the iron
pnictides, previous studies have shown that the large reconstruction
of the band structure dominates the savings of electronic energy,
and would be responsible for the SDW \cite{LXYang, YZhang, MYi},
while there are also suggestions that the SDW might be dominated by
Fermi surface nesting \cite{Dong}. For the iron chalcogenides, a
connection between the electronic structure and the bicollinear
magnetic structure has not been established, except that Fermi
surface nesting has been ruled out \cite{Hasan, NLWang}.  Many
fundamental questions are yet to be addressed for iron
chalcogenides: is there any connection between the electronic
structure and magnetic ordering; and why is it different from the
iron pnictides; what is responsible for the anomalous transport
behaviors in iron chalcogenides? The answers of these questions will
help build a general picture of iron-based superconductors.

In this Letter, we study the electronic structure of a prototypical
parent iron chalcogenide, $\alpha$-Fe$_{1.06}$Te, by angle-resolved
photoemission spectroscopy (ARPES). We found that it is  profoundly
different from those of iron pnictides. The electronic structure of
Fe$_{1.06}$Te is dominated by strong correlation, which induces
incoherent spectra over extended momentum region in the paramagnetic
state. A large square shape of spectral weight unexpectedly appear
around ${\Gamma}$ and extend to $X$ near the Fermi energy ($E_F$).
In the SDW state, with the spectral weight redistribution over a
large energy scale of 0.7~eV, sharp quasiparticle peaks emerge near
$E_F$, indicating reduced spin fluctuations. Through detailed
temperature-dependence studies, we prove that the massive
redistribution of the spectral weight is responsible for the
magnetic transition, unveiling a unique manifestation of SDW on
electronic structure in the presence of strong correlation.

$\alpha$-Fe$_{1.06}$Te single crystals were synthesized following
the method in Ref.~\cite{Transmeth}. Magnetic susceptibility
measurements show that the SDW transition occurs at $T_s$~=~70~K,
accompanied by a structural transition, which is consistent with the
neutron and transport reports \cite{PCDai,NLWang}. ARPES data were
taken with circularly polarized 24~eV photon at the Beamline 9 of
Hiroshima synchrotron radiation center (HiSOR) with a Scienta R4000
electron analyzer. The energy resolution is 10~meV, and angular
resolution is 0.3$^{\circ}$. The sample was cleaved \textit{in
situ}, and measured under ultrahigh vacuum of
$3\times10^{-11}$~{torr}. Aging effects are strictly monitored
during the experiments.

\begin{figure}[t]
\includegraphics[width=8cm]{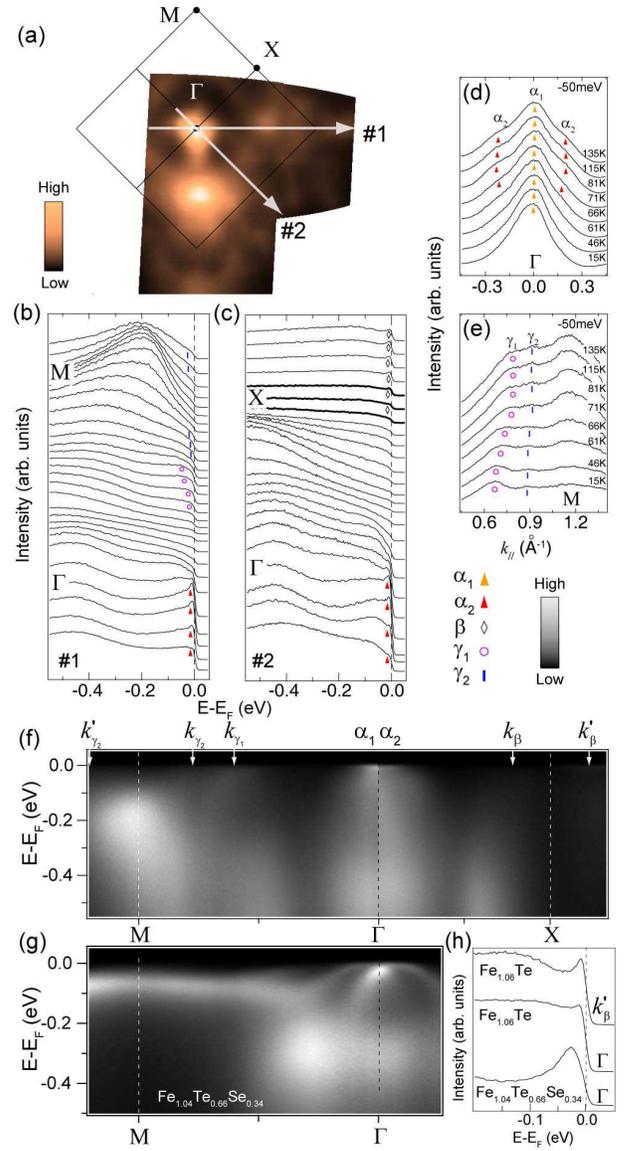}
\caption{(Color online) SDW state electronic structure of
Fe$_{1.06}$Te measured at 15~K. (a) Photoemission intensity
distribution integrated over the energy window of $[E_F-15~meV,
E_F+15~meV]$.  (b)  and (c) The EDCs along cut~\#1 and cut~\#2
respectively.  (d) and (e) are momentum distribution curves (MDCs)
at -50~meV around $\Gamma$ and $M$ at different temperatures
respectively. (f)  The photoemission intensities along the
$M-\Gamma-X$ high symmetry lines. Various Fermi crossings are
denoted by the short arrows. (g) The photoemission intensity of
Fe$_{1.04}$(Te$_{0.66}$Se$_{0.34}$) along $M-\Gamma$. (h) The EDCs
taken at 15~K for Fe$_{1.04}$(Te$_{0.66}$Se$_{0.34}$) at $\Gamma$,
and for Fe$_{1.06}$Te at $\Gamma$ and $k_{\beta}^{'}$ .
}\label{SDWs}
\end{figure}

The electronic structure of Fe$_{1.06}$Te in the paramagnetic state
is shown in Fig.~\ref{normal}. The spectra are characterized by
broad incoherent feature, while the quasiparticle weight is
negligibly small. Thus, Fermi crossings are not well-defined,
analogous to the pseudogap in the cuprates. According to the
spectral-weight distribution near $E_F$ [Figs.~\ref{normal}(a),
~\ref{normal}(c), ~\ref{normal}(d), and ~\ref{normal}(f)], five main
features are labeled as $\alpha_{1,2}$, $\beta$, and $\gamma_{1,2}$.
Their weight distribution is sketched in Fig.~\ref{normal}(b). Note
that, $\alpha_1$ contributes a straight dispersed feature at
$\Gamma$, thus it could not be recognized from EDCs. The features
near $E_F$ around $\Gamma$ and $M$ qualitatively agree with
calculations shown in Fig.~\ref{normal}(g). Thus, the spectral
weight of $\alpha_{1,2}$ and $\gamma_{1,2}$ might be attributed to
the hole and electron pockets respectively, as in
Fe$_{1+y}$Te$_x$Se$_{1-x}$ \cite{ChenFeiPRB}. However, contradicting
to the calculations, a fair amount of spectral weight around  $X$
could be observed near $E_F$ in Fig.~\ref{normal}(d), which is an
extension of the $\beta$ feature.

Many changes occur in the SDW state electronic structure
(Fig.~\ref{SDWs}). Most notably, a dramatic reorganization of the
spectral weight is observed in Fig.~\ref{SDWs}(a), where the weight
suppression around $X$ is particularly strong. Such suppression is
obvious by comparing the  EDCs around $X$ (thick curves) in
Fig.~\ref{SDWs}(c) and Fig.~\ref{normal}(d). Particularly, sharp
quasiparticle peaks appear at $E_F$  around $\Gamma$ and $X$ in
Figs.~\ref{SDWs}(b) and \ref{SDWs}(c). The flat quasiparticle
dispersion and small weight suggest a very low renormalization
factor $Z$. Similar to BaFe$_2$As$_2$, the SDW state of
Fe$_{1.06}$Te is accompanied by band shifts. As shown by MDCs in
Fig.~\ref{SDWs}(d), the distance between two $\alpha_2$ peaks
decreases in the SDW state, indicating a change of dispersion below
$T_s$. The $\gamma_1$ and $\gamma_2$ bands also exhibit an abrupt
momentum shift below $T_s$, illustrating the enlargement of the
electron pockets and band movement around $M$ [Fig.~\ref{SDWs}(e)].

\begin{figure}[t!]
\includegraphics[width=8.5cm]{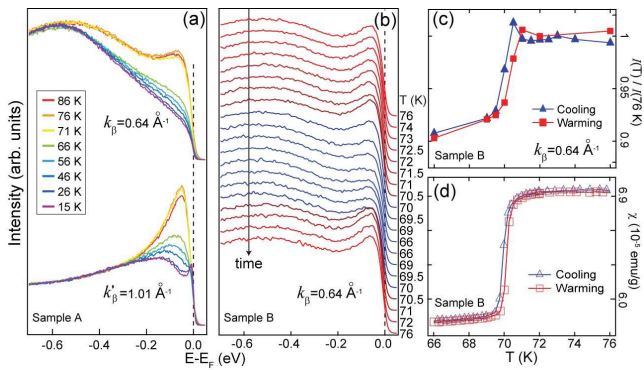}
\caption{(Color online) (a) Detailed temperature dependence of EDCs
at the Fermi crossings of $k_{\beta}$ and $k_{\beta}^{'}$ for sample
A. (b) The EDC at $k_{\beta}$   for sample B in a detailed
temperature cycling experiment near $T_s$. (c)  The integrated
spectral weight over $[E_F-0.7~eV, E_F+0.05~eV]$ as a function of
temperature for data in panel b, data are normalized by the
integrated weight at 75~K. (d) The magnetic susceptibility
hysteresis loop.} \label{TD2}
\end{figure}

The observed incoherent to coherent lineshape evolution explains the
anomalous transport and optical properties of Fe$_{1+y}$Te,
particularly the absence of Drude peak in the paramagnetic state,
and insulator-metal transition as shown in Fig.~\ref{normal}(e)
\cite{NLWang}. Compared with the incoherent weight distribution of
Fe$_{1.06}$Te in Fig.~\ref{SDWs}(f), Fig.~\ref{SDWs}(g) illustrates
the well defined band structure of
Fe$_{1.04}$Te$_{0.66}$Se$_{0.34}$, where the SDW is suppressed by
the heavy Se doping. Since both systems contain similar amount of
interstitial Fe ions, the broad overall lineshape of Fe$_{1.06}$Te
cannot be explained by the magnetic scatting of the excess Fe ions
\cite{NLWang}. Furthermore, as illustrated in Fig.~\ref{SDWs}(h),
the quasiparticle width of Fe$_{1.06}$Te at low temperatures is much
sharper than that of Fe$_{1.04}$Te$_{0.66}$Se$_{0.34}$. This result
together with the narrow transitions in resistivity
[Fig.~\ref{normal}(e)] and magnetic susceptibility
[Fig.~\ref{TD2}(d)] confirm the high quality of Fe$_{1.06}$Te
crystals studied here. Therefore,  the incoherent lineshape over a
large energy scale  should be an \emph{\textbf{intrinsic}} property
of Fe$_{1.06}$Te.

\begin{figure}[t]
\includegraphics[width=7.5cm]{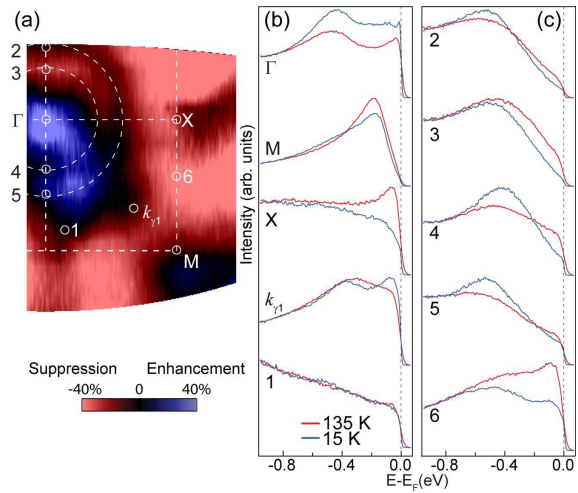}
\caption{(Color online) (a) The difference between the integrated
spectral weight in 135 K and 15 K over the $[E_F-0.7eV,
E_F+0.05eV]$ window for Fe$_{1.06}$Te. (b) and (c) Temperature
dependence of EDCs at various momenta as marked in panel
a.}\label{TD1}
\end{figure}

Due to the low weight of the coherent quasiparticles, and broad
overall lineshape, the dominating effect on the electronic structure
is not the band shift but the substantial spectral weight
redistribution over a large energy scale. Figure~\ref{TD1}(a) plots
the difference between the integrated spectral weight over
$[E_F-0.7~eV, E_F+0.05~eV]$ at 135 and 15~K in the Brillouin zone.
It is clear that spectral weight is suppressed over extended
momentum region, and enhanced around $\Gamma$ at low temperature.
Figs.~\ref{TD1}(b) and \ref{TD1}(c) compare the EDCs at various
representative momenta in the paramagnetic and SDW states. The
enhancement of spectral weight often happens within $[E_F-0.7~eV,
E_F-0.2~eV]$, while the suppression often occurs within
$[E_F-0.4~eV, E_F]$. We note that due to the matrix element effects
caused by different polarization, the difference map is not entirely
symmetric. For example, the high energy part in EDC at momentum 4 is
more prominent than that at momentum 3, although they are symmetric
with respect to $\Gamma$. Overall, a large amount of spectral weight
is transferred from lower binding energies to higher binding
energies, as a result, the electronic energy is significantly
reduced. Such a suppression over a large energy scale is not
relevant to Fermi surface instabilities like nesting. Consistently,
no sign of gap opening is observed in all cases of
Figs.~\ref{TD1}(b) and \ref{TD1}(c). \cite{NLWang,Hasan}. Early DFT
calculations have predicted strong nesting instabilities with
incorrect nesting wavevectors along the $\Gamma-M$ direction
\cite{DJSingh1}. Later on, it has been amended that the excess iron
would significantly alter the electronic structure and produce the
right wavevector\cite{DJSinghdop, LDAdop}. However, this is ruled
out again by the absence of gap observed here.

 Detailed temperature evolution of the spectral-weight redistribution near
$T_s$ is shown in Fig.~\ref{TD2}. The suppression at $k_{\beta}$
occurs abruptly below $T_s$, and saturates at low temperatures
[Fig.~\ref{TD2}(a)]. Furthermore, the temperature cycling experiment
with dense steps around $T_s$ in Fig.~\ref{TD2}(b) gives  a
hysteresis loop in the integrated spectral weight in
Fig.~\ref{TD2}(c), which almost exactly follows the hysteresis loop
in the susceptibility data in Fig.~\ref{TD2}(d) of this first order
transition. This establishes a direct relation between the
suppression and the SDW transition, plus proving that our data
reflect intrinsic and bulk properties. Similar behavior takes place
at $k_{\beta}^{'}$ [Fig.~\ref{TD2}(a)], noting that the difference
between $k_{\beta}$ and $k_{\beta}^{'}$ might be caused by different
$k_z$'s or matrix element effects.

Our observation of the intrinsically incoherent electronic structure
of Fe$_{1.06}$Te and the spectral weight redistribution associated
with SDW transition suggests strong local magnetic fluctuations and
their strong coupling to itinerant electrons. Consequently, carriers
are more localized, causing local moments and insulating transport
behavior \cite{NLWang}, and coherent quasiparticles are destroyed in
the paramagnetic states. However in the SDW state, when the spin
fluctuations are suppressed due to the opening of a spin gap as
demonstrated by inelastic neutron scattering \cite{INS1}, the sharp
quasiparticles emerge. Consistently, it is found that the ordered
moment in FeTe is about 2~$\mu_B$ \cite{PCDai}, much larger than the
0.87~$\mu_B$ in BaFe$_2$As$_2$, or the 0.36~$\mu_B$ in LaOFeAs
\cite{PCDai1111,XHChen122}. Theoretically, the models based on
magnetic exchange interactions between the nearest and next-nearest
neighbor iron moments have successfully explained the bicollinear
magnetic structure in Fe$_{1+y}$Te \cite{JPHuEPL, XTFeTe}. Our data
will be a decisive support, if incoherent electronic structure and
related spectral weight redistribution can be reproduced in these
models.

Furthermore, an early ARPES experiment \cite{Hasan} has shown that
Fe$_{1.05}$Te ($T_s=65$~K)   exhibits an electronic structure close
to that of the non-magnetic Fe$_{1.04}$Te$_{0.66}$Se$_{0.34}$
\cite{ChenFeiPRB}, with more coherent electronic structure. Since
the magnetic order in iron chalcogenides could be strongly
suppressed by just a small amount of Se, excess iron, or pressure
\cite{MaoT,FeTePre,FeTePha}, our samples with higher $T_s$ of 70~K
are in a more strongly ordered state. It is remarkable to observe
that the strong correlation effect is enhanced so dramatically here,
while $T_s$ is just slightly increased. It is sensible to study how
the correlations in iron-based systems are affected by anions
(P/As/Se/Te), doping, and pressure.

Similar behavior has been observed in charge density wave (CDW)
systems like 2H-TaS$_2$, where strong electron-phonon interactions
cause incoherent polaronic spectral lineshape, and spectral weight at the $E_F$ over the entire Brillouin zone. It was found that the massive spectral-weight suppression over a large momentum and energy phase space, instead of Fermi surface nesting,  is responsible for the CDW in 2H-TaS$_2$ and 2H-NbSe$_2$ \cite{Shen1,Shen2}. The analogous mechanism of SDW found here for Fe$_{1.06}$Te indicates the density
waves at the strong coupling limit share a universal theme, which
makes them fundamentally different from the weak interaction
systems.

To summarize, we have carried out a systematic photoemission
investigation of high quality  $\alpha$-Fe$_{1.06}$Te single
crystals. We observed an intrinsically incoherent electronic
structure, and massive spectral weight redistribution that is
responsible for the SDW transition. Our results demonstrate that correlations are probably the strongest in Fe$_{1+y}$Te among all Fe-HTSCs and their parent compounds discovered so far, and reveal universal behaviors of
density waves in the presence of strong interactions.

We thank Dr. Donghui Lu for helpful discussions. This work was
supported by the NSFC, MOE, MOST (National Basic Research Program
No.~2006CB921300), and STCSM of China.

\end{document}